# Evolution of safety factor profiles in sawteeth

W. Zhang[1], Z. W. Ma[1,a)] and H. W. Zhang[1]

[1]Institute for Fusion Theory and Simulation, Department of Physics, Zhejiang University, Hangzhou 310027, China

**Abstract:** Two different definitions of the safety factor are applied to investigate the evolution of the safety factor profile during normal sawteeth, the stationary state, and the incomplete reconnection. It is found that the safety factor profiles from the old definition are sometimes inconsistent with the Poincare plots of the magnetic field during sawteeth. The old safety factor always indicates that the safety factor around the magnetic axis is flattened and equal to 1.0 with the development of the kink instability. However, the Poincare plots of the magnetic field lines indicate that the topology of the magnetic field around the magnetic axis has not been changed. To solve the inconsistency, we propose a new definition of the safety factor, in which the poloidal angle relative to the new twisted magnetic axis is used instead of the poloidal angle to the original axis. With the new definition, the safety factor profiles are consistent with Poincare plots of the magnetic field. We also find that the safety factor profiles are significantly different from the two different q definitions. With the new q definition, the safety factor at the magnetic axis $q_0$ remains unchanged in almost the entire period of a sawtooth and jumps up to 1.0 near the end during normal sawteeth; in the non-axisymmetric equilibrium, $q_0$ is still far below 1.0; $q_0$ remains its initial value throughout the incomplete reconnection.

E-mail: zwma@zju.edu.cn

## I. Introduction

Sawteeth are common phenomena for magnetically confined fusion device, whose central safety factor falls below one. [1] During sawteeth, the central plasma pressure periodically crashes after a slow rise. Sawteeth can not only flatten center plasma temperature but also trigger neo-classical tearing modes in nearby resonant surfaces[2], which results in a significant reduction of energy confinement.

Since the sawteeth are deleterious for Tokamak operations, many efforts were taken to understand the mechanism of sawteeth.[3-5] However, after more than 40 years, two fundamental points of sawteeth is still unknown, i.e., the mechanism of the fast pressure crash and whether the magnetic reconnection is complete or incomplete during the crash. For the first problem, there are several candidates, i. e. the Hall effect,[6] the stochasticity of the magnetic field[7], and the pressure-driven instabilities.[8] For the second problem, we are even unable to know whether the incomplete reconnection has actually occurred since the q profile evolution from different Tokamaks is significantly different. In TFTR[9] and ASDEX-U[10], the safety factor of the magnetic axis $q_0$ remains almost unchanged during sawteeth. However, in other experiments, $q_0$ goes above one after the crash. Therefore, the calculation of q profiles is of great importance to understand the physical mechanism of the sawteeth, especially for incomplete reconnection.

The safety factor is defined as $q = \frac{\Delta\varphi}{\Delta\theta}$ to reflect the helicity of a magnetic field line, where $\Delta\varphi$ and $\Delta\theta$ are the changes of the toroidal and poloidal angles along the magnetic field line based on the initial untwisting magnetic axis. However, if the magnetic axis is twisted due to kink instabilities, the magnetic field lines not only twist with its helicity but also have to wind around the twisted magnetic axis. If we still use such a definition, the q profile will totally be misleading and deceptive. For example, if the 1/1 kink instability is well developed, the magnetic axis will be twisted, and its helicity is m/n=1/1. Assuming the twisted magnetic axis at a toroidal plane locates at $(r_A, \theta_{A0} - \varphi, \varphi)$, where $r_A$ is the distance between the locations of the new and initial magnetic axis, $\theta_{A0}$ is the poloidal angle of the new twisted magnetic axis at $\varphi=0$. Then, the twisted magnetic axis locates at

$$R_A = R_{A0} + r_A \cos(\theta_{A0} - \varphi), \tag{1}$$

$$Z_A = Z_{A0} + r_A \sin(\theta_{A0} - \varphi), \tag{2}$$

where ($R_{A0}, Z_{A0}=0$) is the position of the untwisted magnetic axis at $\varphi=0$. Magnetic field lines in the region $\sqrt{(R-R_A)^2 + (Z-Z_A)^2} < r_A$ now wind around the twisted magnetic axis while magnetic field lines in other regions are not affected. For a magnetic field line starts from $(R_A+\tilde{r},0,0)$ and $\tilde{r} < r_A$, its helicity is $q_F$, the location of the magnetic field line at each plane will be

$$R(\varphi) = R_{A0} + r_A \cos(\theta_{A0} - \varphi) + \tilde{r}\cos(-\varphi/q_F) \tag{3}$$

$$Z(\varphi) = r_A \sin(\theta_{A0} - \varphi) + \tilde{r}\sin(-\varphi/q_F) \tag{4}$$

We name the old safety factor as,

$$q_{old} = \frac{\Delta\varphi}{\Delta\theta} = \frac{\Delta\varphi/(2\pi)}{N_{z=0}/2} = \frac{\Delta\varphi}{\pi N_{z=0}}, \tag{5}$$

where $N_{z=0}$ is the time for a magnetic field line crossing the $Z=0$ plane. From $Z(\varphi) = 0$, we get

$$r_A \sin(\theta_{A0} - \varphi) + \tilde{r}\sin(-\varphi/q_F) = 0 \tag{6}$$

Since $\tilde{r} < r_A$, then

$$\Delta\varphi = 2[\frac{N_{z=0}}{2}]\pi + \theta_{A0} + \arcsin[\frac{\tilde{r}}{r_A}\sin(-\Delta\varphi/q_F)] \tag{7}$$

or

$$\Delta\varphi = (2[\frac{N_{z=0}}{2}]+1)\pi + \theta_{A0} - \arcsin[\frac{\tilde{r}}{r_A}\sin(-\Delta\varphi/q_F)] \tag{8}$$

To get an accurate safety factor, we have $N_{z=0} \gg 1$ (typically $N_{z=0} \sim 1000$). Thus,

$$\Delta\varphi \approx \pi N_{z=0} \tag{9}$$

And the old safety factor will be

$$q_{old} = \frac{\Delta\varphi}{\pi N_{z=0}} \equiv 1 \quad (10)$$

Equation (10) indicates that the old definition of the safety factor has not considered that the magnetic axis is twisted due to the kink instability, and all magnetic field lines in the region must wind around the twisted magnetic axis. As a result, no matter what the helicity of the magnetic field it is, the old definition of q always 'proves' that the profile of the old safety factor around the magnetic axis ($\tilde{r} < r_A$) is flattened, and they are equal to unit. It should be noted that the helicity or topology of a magnetic field line should remain unchanged unless the field line is reconnected. If the magnetic field lines in the region ($\tilde{r} < r_A$) have not been reconnected, the safety factor should remain its initial value. As we can see, the safety factor from Equation (1.10) fails to reflect the helicity of the magnetic field, and it always gives a wrong q value. The reason is that, in this region, the old definition of q has not taken into the influence of the twisted magnetic axis.

In the region $\sqrt{(R-R_A)^2+(Z-Z_A)} > r_A$, since the magnetic field line is not affected by the twisted magnetic axis, the old q definition gives a right value, i.e.,

$$q_{old} \equiv q_F \quad (11)$$

Note that magnetic field lines in the m/n=1/1 island do not wind around the new magnetic axis, and the safety factor in the island is not affected. From the above discussion, the old safety factor will always give a wrong value in the region near the twisted magnetic axis. That is the reason why the q profiles sometimes are inconsistent with the Poincare plots of magnetic field lines during sawteeth.

**II. A new method for the safety factor calculation**

Since the main problem is resulted from the influence of the twisted magnetic axis, we introduce the new poloidal angle $\theta'=\theta-\theta_A$, where $\theta$ and $\theta_A$ are the poloidal angles of a magnetic field line and the twisted magnetic axis. Thus, the new poloidal angle $\theta'=\theta-\theta_A$ of the magnetic field line is the poloidal angle, which is

4 / 18

relative to the twisted magnetic axis. In the region $\sqrt{(R-R_A)^2+(Z-Z_A)} < r_A$, the relative position of the magnetic field line to the twisted magnetic axis is

$$R_{new}(\varphi) = \tilde{r}\cos(-\varphi/q_F) \tag{12}$$

$$Z_{new}(\varphi) = \tilde{r}\sin(-\varphi/q_F) \tag{13}$$

Similarly, the new safety factor

$$q_{new} = \frac{\Delta\varphi}{\Delta\theta'} = \frac{\Delta\varphi/(2\pi)}{N_{z_{new}=0}/2} = \frac{\Delta\varphi}{\pi N_{z_{new}=0}}, \tag{14}$$

where $N_{z_{new}=0}$ is the time for the magnetic field line crossing the $Z_{new}=0$ plane. From $Z_{new}(\varphi)=0$,

$$\tilde{r}\sin(-\varphi/q_F) = 0, \tag{15}$$

$$\Delta\varphi = q_F \pi N_{z_{new}=0}, \tag{16}$$

Then,

$$q_{new} = \frac{\Delta\varphi}{\pi N_{z_{new}=0}} \equiv q_F \tag{17}$$

Equation (1.16) indicates that the new safety factor in the region $\sqrt{(R-R_A)^2+(Z-Z_A)} < r_A$ can successfully solve the problem resulted from the influence of the twisted magnetic axis.

Now we calculate the safety factor in the region with $\sqrt{(R-R_A)^2+(Z-Z_A)} > r_A$. Since magnetic field lines are not affected by the twisted magnetic axis,

$$R_{new}(\varphi) = R_{A0} + \tilde{r}\cos(-\varphi/q_F) - r_A\cos(\theta_{A0}-\varphi) \tag{18}$$

$$Z_{new}(\varphi) = \tilde{r}\sin(-\varphi/q_F) - r_A\sin(\theta_{A0}-\varphi) \tag{19}$$

From $Z_{new}(\varphi)=0$,

$$\tilde{r}\sin(-\varphi/q_F) - r_A\sin(\theta_{A0}-\varphi) = 0 \tag{20}$$

Since $\tilde{r} > r_A$,

$$\Delta\varphi = q_F \{2[\frac{N_{z_{new}=0}}{2}]\pi - \arcsin[\frac{r_A}{\tilde{r}}\sin(\theta_{A0} - \Delta\varphi)]\}, \quad (21)$$

or

$$\Delta\varphi = q_F \{(2[\frac{N_{z_{new}=0}}{2}]+1)\pi + \arcsin[\frac{r_A}{\tilde{r}}\sin(\theta_{A0} - \Delta\varphi)]\} \quad (22)$$

thus

$$q_{new} = \frac{\Delta\varphi}{\pi N_{z_{new}=0}} \equiv q_F. \quad (23)$$

Equations (17) and (23) indicate that, when the magnetic axis is twisted by the kink instabilities, we can obtain the right safety factor by using the poloidal angle relative to the twisted magnetic axis during q calculation. It is also should be noted that the new safety factor calculation method can be applied in the whole region, not only in the region where the magnetic field lines have to wind around the twisted magnetic axis.

### III. Simulation results

All the simulations in the present paper are carried out with the CLT code. [11] Since the purpose of the simulations is to verify the accuracy of the new safety factor, we do not repeat the details of the CLT code. Similar simulation results could be found in our previous studies (W. Zhang et al. to be published).

#### i. Normal sawteeth

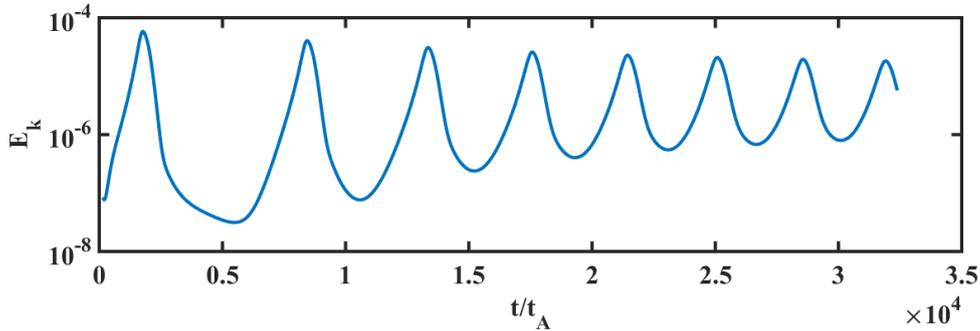

Figure 1. The kinetic energy evolution of normal sawteeth.

The parameters used in this subsection are given as follows: the plasma beta $\beta_0 \sim 2.4\%$, the resistivity $\eta = 2.5\times10^{-6}$, the diffusion coefficient $D = 1.0\times10^{-5}$, the viscosity $\nu=1.0\times10^{-4}$, the perpendicular and parallel thermal conductivities $\kappa_\perp = 2.0\times10^{-5}$ and $\kappa_\parallel = 5\times10^{-2}$, respectively. The kinetic energy evolution during normal sawteeth is shown in Figure 1. The Poincare plots of magnetic field lines and the q profiles at four typical moments in the first cycle ($t=0t_A$, $t=1243t_A$, $t=1597t_A$, and $t=1775t_A$) are shown in Figure 2. As shown in Figure 2e and 2h, the old and new definition give the same profiles when the amplitude of the kink mode is small or magnetic reconnection finishes. However, when the m/n=1/1 magnetic island appears, the old and new safety factors at the magnetic axis are significantly different. The old safety factor is 1.0 at $t=1243t_A$ and $t=1597t_A$ as shown in Figure 2f and 2g. However, the new safety factor remains $q_0=0.7$, which is its initial value.

It should be noted that the helicity (or topology) of magnetic field lines should remain unchanged unless the magnetic field line is reconnected. Therefore, the safety factor at the magnetic axis should remain its initial value at $t=1243t_A$ and $t=1597t_A$. However, the old definition indicates that $q_0=1.0$, and the q profile becomes flattened around the magnetic axis, which is obviously wrong. As pointed out in the introduction, the wrong results come from the influence of the twisted magnetic axis. Taking into account the influence of the twisted magnetic axis, $q_0$ remains its initial value even when the nearby magnetic field lines start to reconnect. The contour plots of the old and new safety factors at $t=1597t_A$ are shown in Figure 3. The profile of the new safety factor agrees well with the Poincare plots of the magnetic field lines, and the old safety factor profile is far from the Poincare plots.

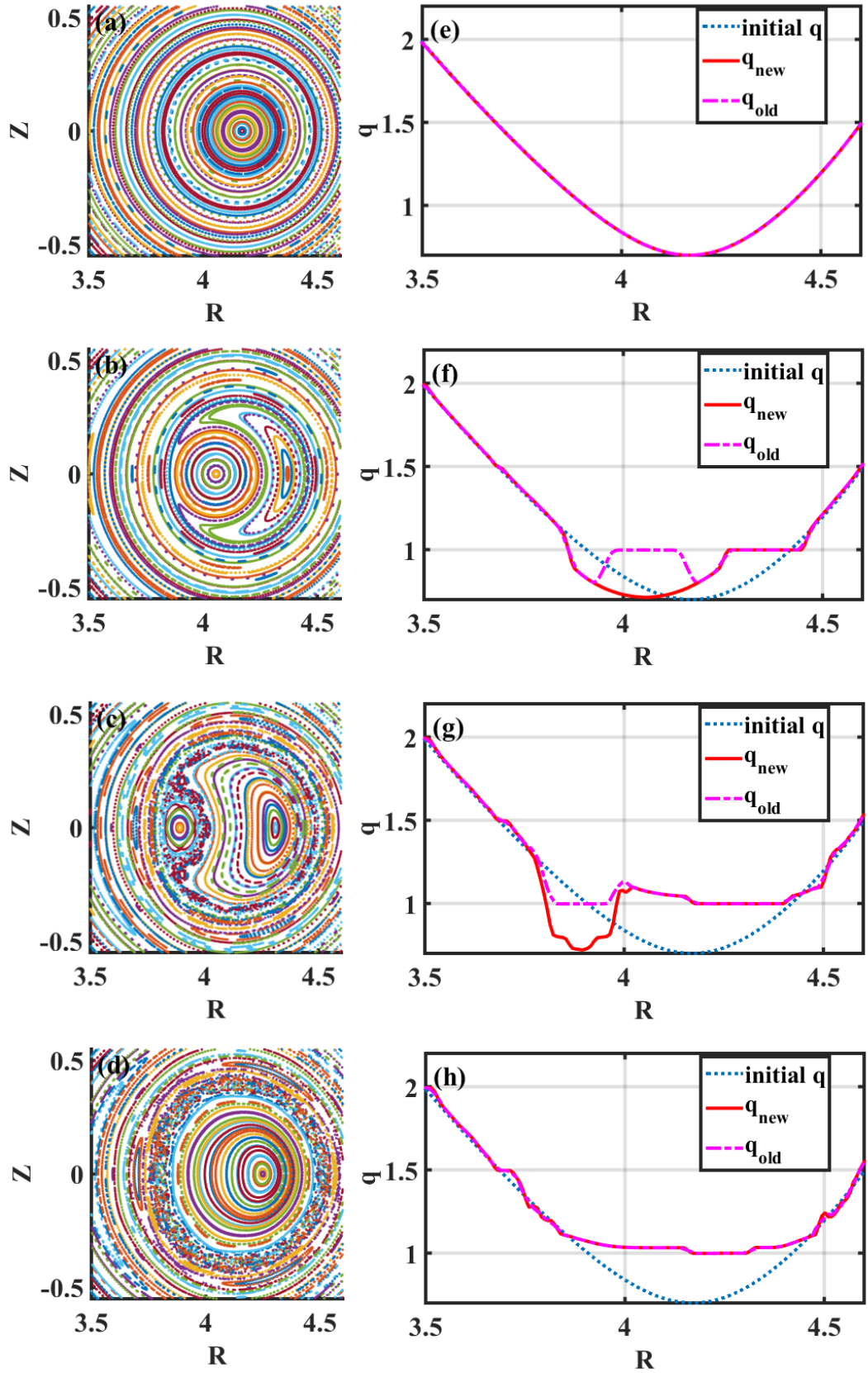

Figure 2 The Poincare plots of the magnetic field and the profiles of the safety factor at $t=0t_A$, $t=1243t_A$, $t=1597t_A$, and $t=1775t_A$. $q_{old}$ is the safety factor with the old poloidal angle that is still defined based on the untwisting magnetic axis, and $q_{new}$ is the safety factor with the new poloidal angle that is redefined based on the twisting magnetic axis.

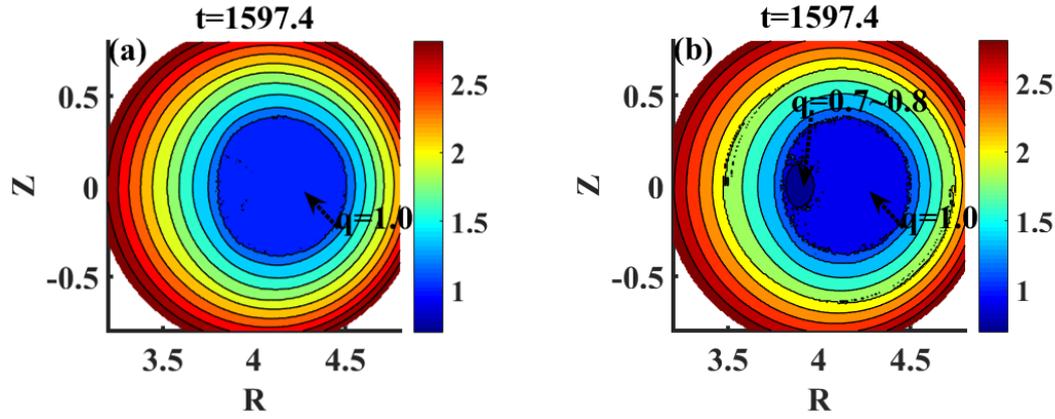

Figure 3 The contour plots of the old (a) and new (b) safety factors at $t = 1597 t_A$.

It should be noted that, even at the moment ($t = 1597 t_A$), when the original core almost disappears, the safety factor at the magnetic axis still keeps $q_0 \sim 0.7$. It indicates that the safety factor at the magnetic axis almost keeps unchanged during the reconnection (Figure 2g), and suddenly jumps up to 1.0 until the magnetic flux reconnection finishes (Figure 2h). From Wesson's theory [5], we know that the interchange instability (or quasi-interchange instability[8]) can only occur when the magnetic shear $s = \dfrac{rq'}{q}$ becomes much smaller. However, as shown in Figure 2f and 2g, the magnetic shear at the X-point becomes larger instead of smaller, which implies that the quasi-interchange instability should not be responsible for the fast pressure crash during the sawteeth in Tokamaks unless the initial safety factor profiles around the magnetic axis are flattened and close to unit.

The evolutions of the safety factor at the magnetic axis with four different definitions are shown in Figure 4. Firstly, the minimum $q_{new}$, which is the new safety factor at the new magnetic axis, keeps almost unchanged for a long time and suddenly jumps up to 1.0 at the end of the reconnection process. The minimum $q_{old}$, which was wrongly regarded as the safety factor at the magnetic axis, gradually rises to 1.0 before the reconnection finishes (i. e. Figure 2b, 2c, 2f, and 2g). Moreover, if one uses $q_{old}(0)$, which is the old safety factor that located at the original axis. The axis safety factor will keep larger than or equal to 1.0 during the sawteeth.

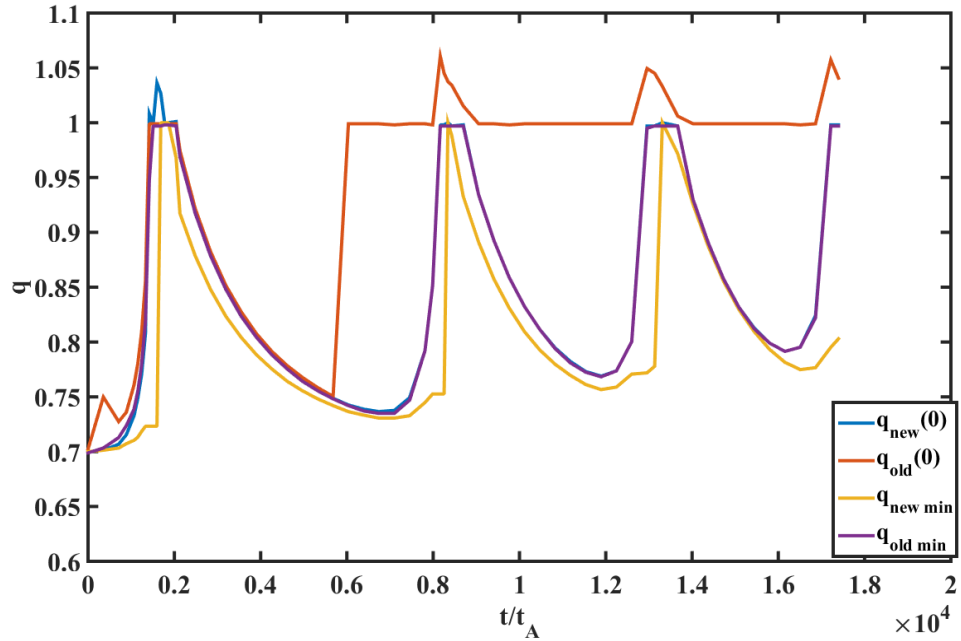

Figure 4 The evolutions of the safety factor at the magnetic axis for four different calculations. $q_{new}(0)$ indicates the new safety factor located at the original magnetic axis. $q_{old}(0)$ indicates the old safety factor located at the original magnetic axis. $q_{new\_min}$ represents the minimum new safety factor along X=0, which is the new safety factor at the new magnetic axis. $q_{old\_min}$ represents the minimum old safety factor along X=0.

### ii. Stationary state

Recently a non-axisymmetric stationary state that is related to sawteeth has been reported in many experiments[12-14]. In those papers, the magnetic field and the stream function both have the helicity of m/n=1/1 at the stationary state. It also has been reported that the safety factor in the core region becomes flattened and is about 1.0. However, if the new q calculation method is applied, the safety factor profile at the stationary state is not entirely flattened and is still smaller than 1.0. We could illustrate this by carrying out similar simulations.

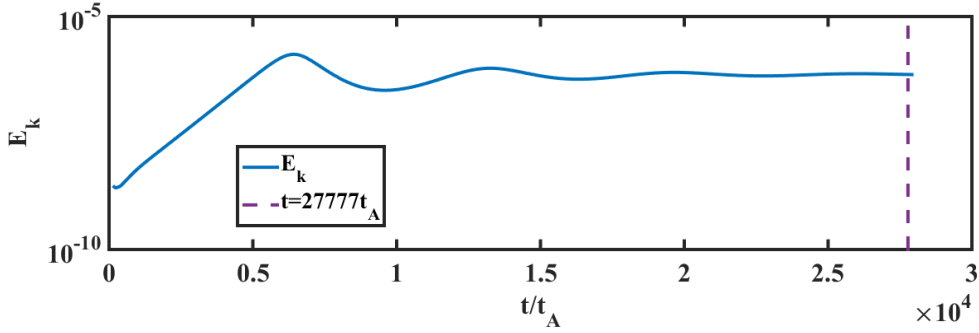

Figure 5 The kinetic energy evolution of the stationary steady of sawteeth with high viscosity.

The parameters used in this subsection are given as follows: the plasma beta $\beta_0 \sim 2.4\%$, the resistivity $\eta = 2.5 \times 10^{-6}$, the diffusion coefficient $D = 1.0 \times 10^{-5}$, the perpendicular and parallel thermal conductivities $\kappa_\perp = 2.0 \times 10^{-5}$ and $\kappa_\parallel = 5 \times 10^{-2}$, and the viscosity $\nu = 1.0 \times 10^{-3}$, respectively. The kinetic energy evolution of the stationary state with high viscosity is shown in Figure 5. The Poincare plot of magnetic field lines at the stationary state is typically like Figure 6(a). As shown in Figure 6 (b), the old safety factor indicates that the safety factor around the magnetic axis is totally flattened and just above 1.0, while the new safety factor indicates that its safety factor still remains below 1.0, $q_{new}=0.86$. The contour plots of the new and old safety factors at $t = 27777 t_A$ are shown in Figure 7. It is evident that the contour plot of the new safety factor is consistent with the Poincare plot of the magnetic field, while the old safety factor profile is not.

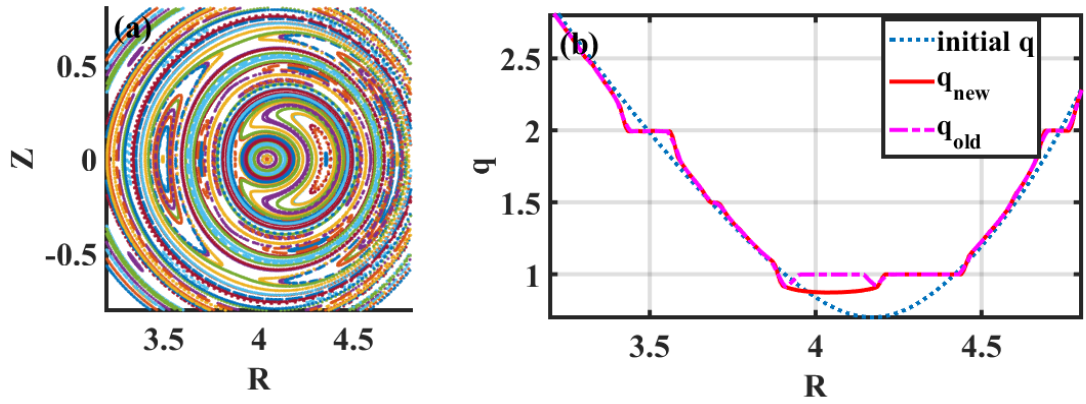

Figure 6 (a) The Poincare plot and (b) the profiles for the old and new safety factors at the stationary state ($t = 27777t_A$).

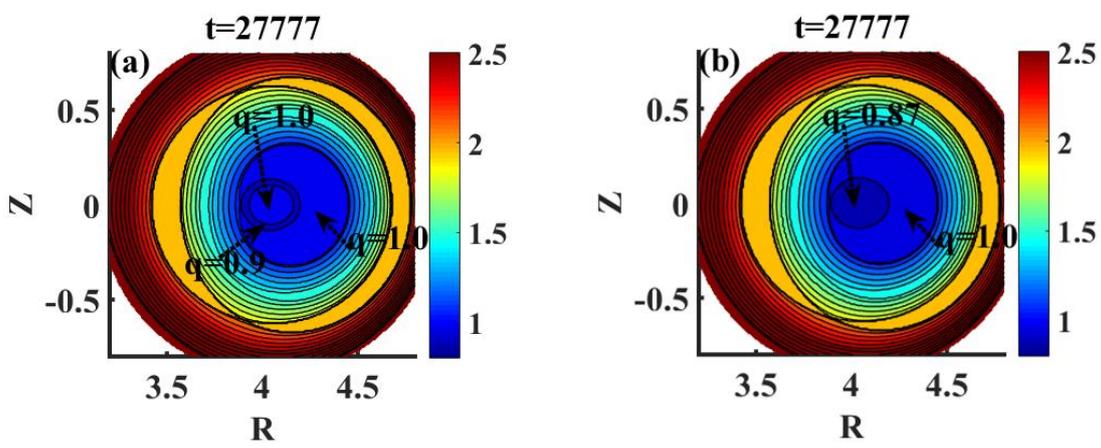

Figure 7 The contour plots of the (a) old and new (b) safety factors at $t = 27777t_A$.

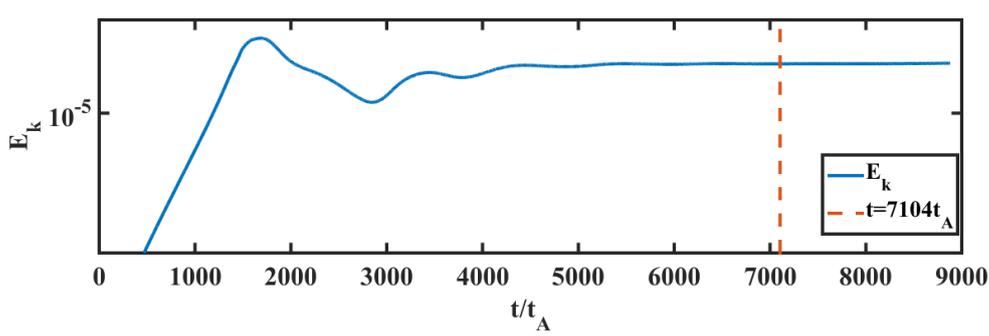

Figure 8 The kinetic energy evolution of the stationary state of sawteeth with low viscosity.

The evolution of kinetic energy with low viscosity ($v=6.0\times10^{-6}$) is shown in Figure 8. The system could also achieve the stationary state with a large m/n=1/1 magnetic island in the present case (Figure 9a) rather than with a small m/n=1/1 island with high viscosity. The corresponding profiles of the old and new safety factors are shown in Figure 9b. The real safety factor at the magnetic axis is 0.9387, which is still below 1.0, which could also be seen from the contour plot of the safety factor (Figure 10 (b)).

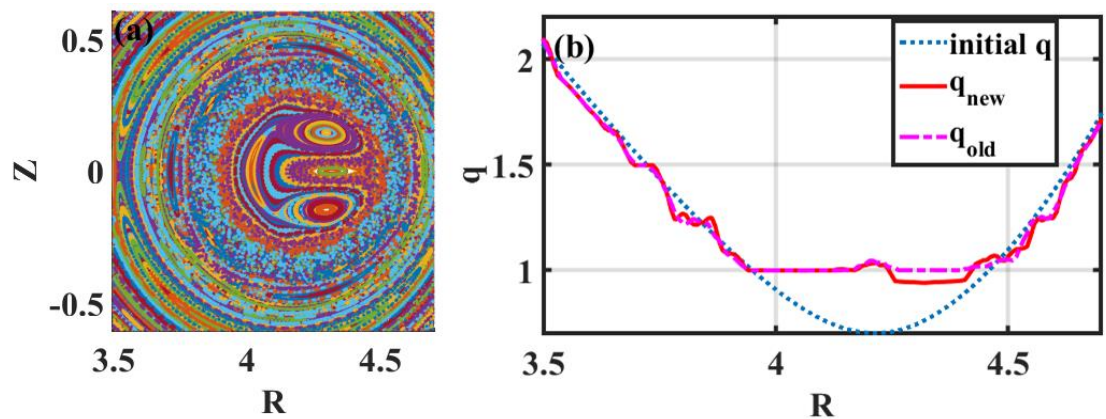

Figure 9 (a) The Poincare plot and (b) the profiles of the old and new safety factors at the stationary state ($t = 7104 t_A$).

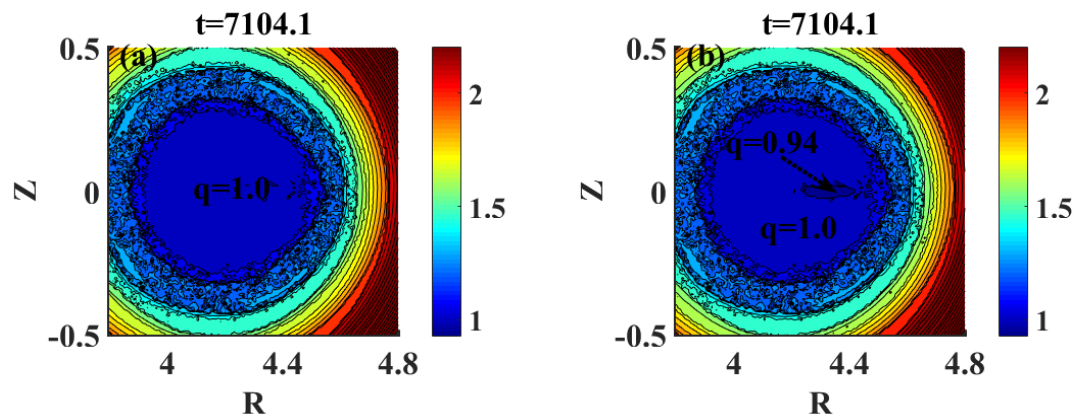

Figure 10 The contour plots for the (a) old and (b) new safety factors at $t = 7104 t_A$.

### iii. incomplete reconnection

Several studies [15, 16] have reported that magnetic reconnection during sawteeth could be incomplete due to plasmoid instabilities. It is interesting and important to calculate the evolution of the safety factor. The parameters used in this subsection are given as follows: the plasma beta $\beta_0 \sim 0$, the resistivity $\eta = 1.0 \times 10^{-7}$, the diffusion coefficient $D=1.0\times10^{-4}$, the perpendicular and parallel thermal conductivities $\kappa_{\perp}=3.0\times10^{-6}$ and $\kappa_{\parallel}=5\times10^{-2}$, and the viscosity $\nu=1.0\times10^{-8}$, respectively. The Poincare plot of magnetic field lines and the corresponding q profiles of the old and new safety factors at four typical moments (at $t=0t_A$, $t=3423t_A$, $t=4336t_A$, and $t=5324t_A$) are shown in Figure 11. The system is unstable for the resistive-kink mode since the initial safety factor at the magnetic axis is 0.9 (Figure 11 e). During the development of the m/n=1/1 resistive-kink mode(Figure 11b), the current sheet near the X-point becomes thinner and thinner. When the current sheet thickness decreases below a critical value, a secondary tearing instability will be triggered, and plasmoids form near the original X-point (Figure 11c). The secondary islands finally merge and form a large secondary island, which prevents the resistive-kink mode from further growing up and then finally results in an incomplete reconnection (Figure 11d). As shown in Figure 11e~11f, the profiles of the new and old safety factors are the same except the region near the magnetic axis. The old safety factor indicates that the safety factor is flattened and becomes equal to 1.0. However, as shown in the Poincare plot, magnetic reconnection only occurs on the q=1 resonant surface instead of occurring around the magnetic axis. Therefore, the safety factor at the magnetic axis should remain unchanged during the incomplete reconnection. From the profile of the new safety factor, the safety factor at the magnetic axis indeed remains 0.9 throughout the simulation, and the safety factor profile only becomes flattened in the two 1/1 magnetic islands. The contour plots of the old and new safety factors at $t=4336t_A$ are shown in Figure 12 a and b. It is clear that the contour plots of the new safety factor agree well with the Poincare plots, while the results from the old safety factor do not.

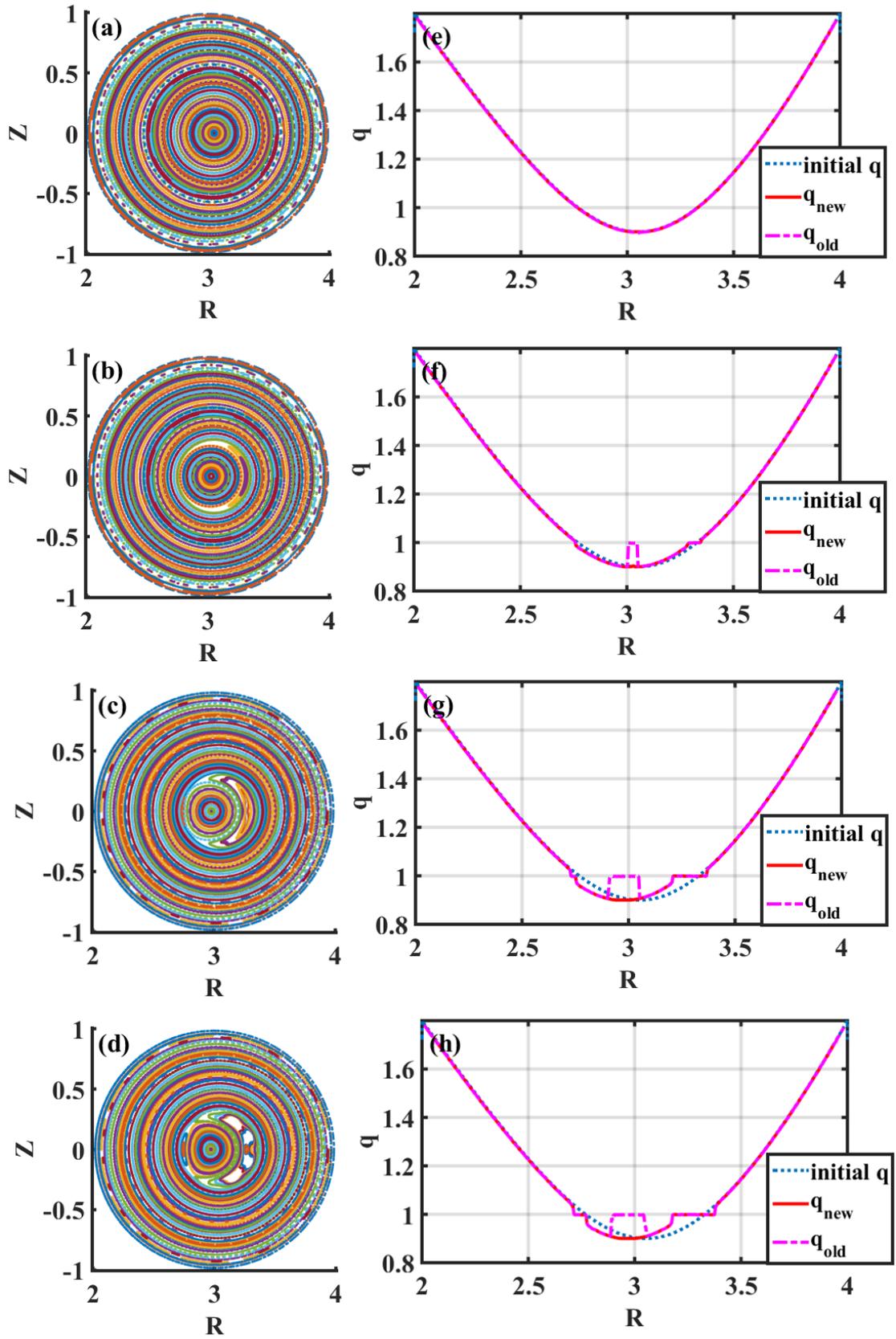

Figure 11 The Poincare plots of the magnetic field and the corresponding profiles of the old and new safety factors at $t=0t_A$, $t=3423t_A$, $t=4336t_A$, and $t=5324t_A$.

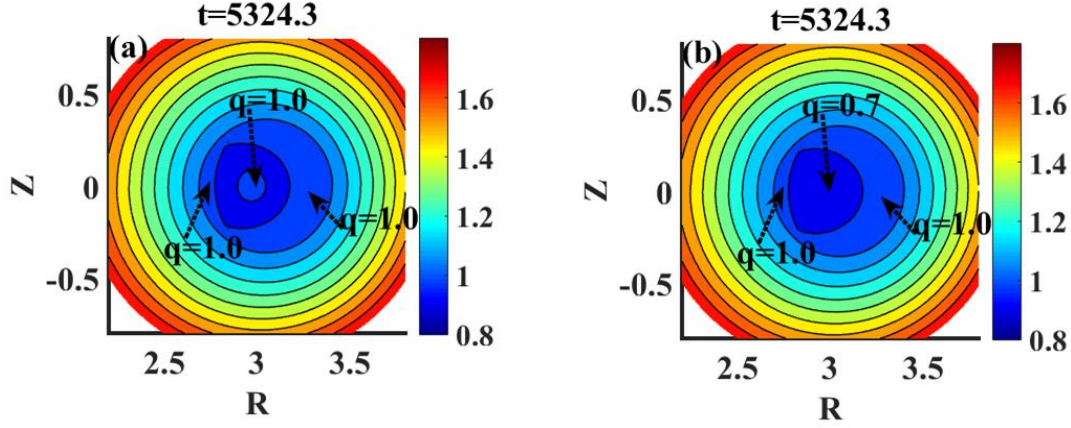

Figure 12 The contour plots of the (a) old and new (b) safety factors at $t = 5324 t_A$.

## IV. Summary and discussion

The safety factors from two different definitions are adopted to investigate the evolution of the safety factor profile during normal sawteeth, the stationary state, and the incomplete reconnection. We find that the safety factor profiles from the old definition are sometimes inconsistent with the Poincare plots of the magnetic field. The old safety factor always indicates that the safety factor around the magnetic axis is flattened and equal to 1.0 with the development of the kink instability.

A new definition of safety factors is proposed, in which the poloidal angle relative to the new twisted magnetic axis is used instead of the poloidal angle relative to the original magnetic axis. It is found that the new safety factor agrees well with the Poincare plots of the magnetic field for all kinds of sawteeth. The new safety factor at the magnetic axis remains unchanged for quite a long time of a sawtooth cycle and then quickly jumps up to 1.0 near the end of magnetic reconnection. For the new safety factor at the magnetic axis, there is no slow ramping phase. Instead, only a sudden transition phase (i.e. $q_0$ jumps up from the initial safety factor to 1.0) that occurs near the end of magnetic reconnection for the normal sawteeth. It should also be noted that the old safety factor at the location of the original magnetic axis always gives $q_0 = 1.0$ during the entire period of sawteeth.

With the old definition, the safety factor at the magnetic axis is flattened and is about 1.0 when the system achieves the steady-state. However, the Poincare plots of the magnetic field indicate that the safety factor at the magnetic axis should not be 1.0. The inconsistency is resulted from the influence of the twisted magnetic axis. With the new definition of q, the safety factor at the stationary state still remains below 1.0, instead of a flattened profile with $q=1.0$ around the magnetic axis. It is evident that only the q profile from the new definition is consistent with the Poincare plots of the magnetic field.

For the incomplete reconnection case, the helicity of the magnetic field near the magnetic axis should remain unchanged during sawteeth. Nevertheless, the old safety factor definition indicates the safety factor at the magnetic axis becomes flattened and equal to 1.0, which is apparently incorrect. From the new definition of safety factors, the safety factor indeed remains unchanged during sawteeth, which is consistent with the experimental observations.

**Acknowledgment**

Dr. Wei Zhang would like to thank Prof. Guoyong Fu and Prof. Francesco Porcelli for their helpful comments. This work is supported by the National Natural Science Foundation of China under Grant No. 11775188 and 11835010, the Special Project on High-performance Computing under the National Key R&D Program of China No. 2016YFB0200603, Fundamental Research Fund for Chinese Central Universities.